\documentclass[twocolumn,showpacs,preprintnumbers,amsmath,amssymb]{revtex4}
  \usepackage{multirow}

\usepackage{graphicx}
\usepackage{dcolumn}
\usepackage{bm}

\begin{document}

\preprint{APS/123-QED}

\title{Origins of Taylor's power law for fluctuation scaling in complex
systems}

\author{Agata Fronczak}\email{agatka@if.pw.edu.pl}
\author{Piotr Fronczak}\email{fronczak@if.pw.edu.pl}

\affiliation{%
Faculty of Physics, Warsaw University of Technology,\\
Koszykowa 75, PL-00-662 Warsaw, Poland
}%

\date{\today}

\begin{abstract}
Taylor's fluctuation scaling (FS) has been observed in many natural and man-made systems revealing an amazing universality of the law. Here we give a reliable explanation for the origins and abundance of Taylor's FS in different complex systems. The universality of our approach is validated against real world data ranging from bird and insect populations through human chromosomes and traffic intensity in transportation networks to stock market dynamics. Using fundamental principles of statistical physics (both equilibrium and non-equilibrium) we prove that Taylor's law results from the well-defined number of states (NoS) of a system characterized by the same value of a macroscopic parameter (i.e., the number of birds observed in a given area, traffic intensity measured as a number of cars passing trough a given observation point or daily activity in the stock market measured in millions of dollars).
\end{abstract}

\maketitle

\section{Introduction and motivation}

In ecology, Taylor's power law~\cite{1961NatureTaylor} (or the law of the mean) states that the mean, $\langle N\rangle$, and the variance, $\sigma^2_N=\langle N^2\rangle-\langle N\rangle^2$, characterizing the number of population representatives are related by power law,
\begin{equation}\label{TaylorLaw}
\langle N^2\rangle-\langle N\rangle^2=a\langle N\rangle^b,
\end{equation}
with the characteristic exponent, $b$, describing effects of heterogeneity in spatial or temporal patterns of the frequency distribution. The value of $b$ is usually in the range of $1$ to $3$. For comparison with the Poisson distribution, $b=1$, the parameter $b>1$ corresponds to clustering (aggregation), whereas $b<1$ may be interpreted as ordering.

As we have already stated, the FS described by Eq.~(\ref{TaylorLaw}) has been noted in a variety of natural and man-made systems, and the universality of Taylor's law is now widely recognized~\cite{2002PTBSBrown,2002ApplEcolHolt,2004PRLMenezes1, 2008AdvPhysEisler}. To emphasize that the generality of our approach to this law is commensurate with the universality of the law itself, we do not concentrate on a specific system in this paper. Later in the text, the macroscopic quantity $N$ simultaneously stands for the number of birds observed in a given area, the number of cavities and corn borers found on one plant, the number of gene structures located in equal-sized non-overlapping bins that span the whole chromosome, the daily traded value of a given stock at the New York Stock Exchange (NYSE), and the number of cars passing through a certain observation point in a given time period.

When the data over which the averages (\ref{TaylorLaw}) are taken have a temporal structure (like in the case of traffic), the Taylor's law is called {\it temporal} FS, otherwise (like in the case of chromosomes) we term it {\it ensemble} FS \cite{2008AdvPhysEisler}. We show, that given our approach one does not need to invoke any stochastic models to explain phenomena such as aggregation effects in different populations, and traffic jams. Regardless of character of the scaling (ensemble or temporal) the collective phenomena manifesting themselves by Taylor's law simply result from properties of the phase space underlying the considered systems.

The only assumption made concerning the systems obeying the Taylor's law is that they are either in equilibrium or in non-equilibrium steady states. The equilibrium formalism described in the paper is suited for systems with ensemble FS that are characterized by stable, time-independent probability distributions. Systems with temporal FS described by  time series with a stable (i.e. constant in time) mean value and variance, are treated as non-equilibrium systems in steady states. In the paper, the case of ensemble FS is represented by the data on spatial distribution of larval populations of the European corn borer and the data on spatial distribution of gene structures on human chromosome 7, while the case of temporal FS is represented by traffic in Minnesota and the steady-state dynamics of North American avifauna and NYSE.

\section{Data description}\label{Data}

\begin{figure*}
\includegraphics[width=\textwidth]{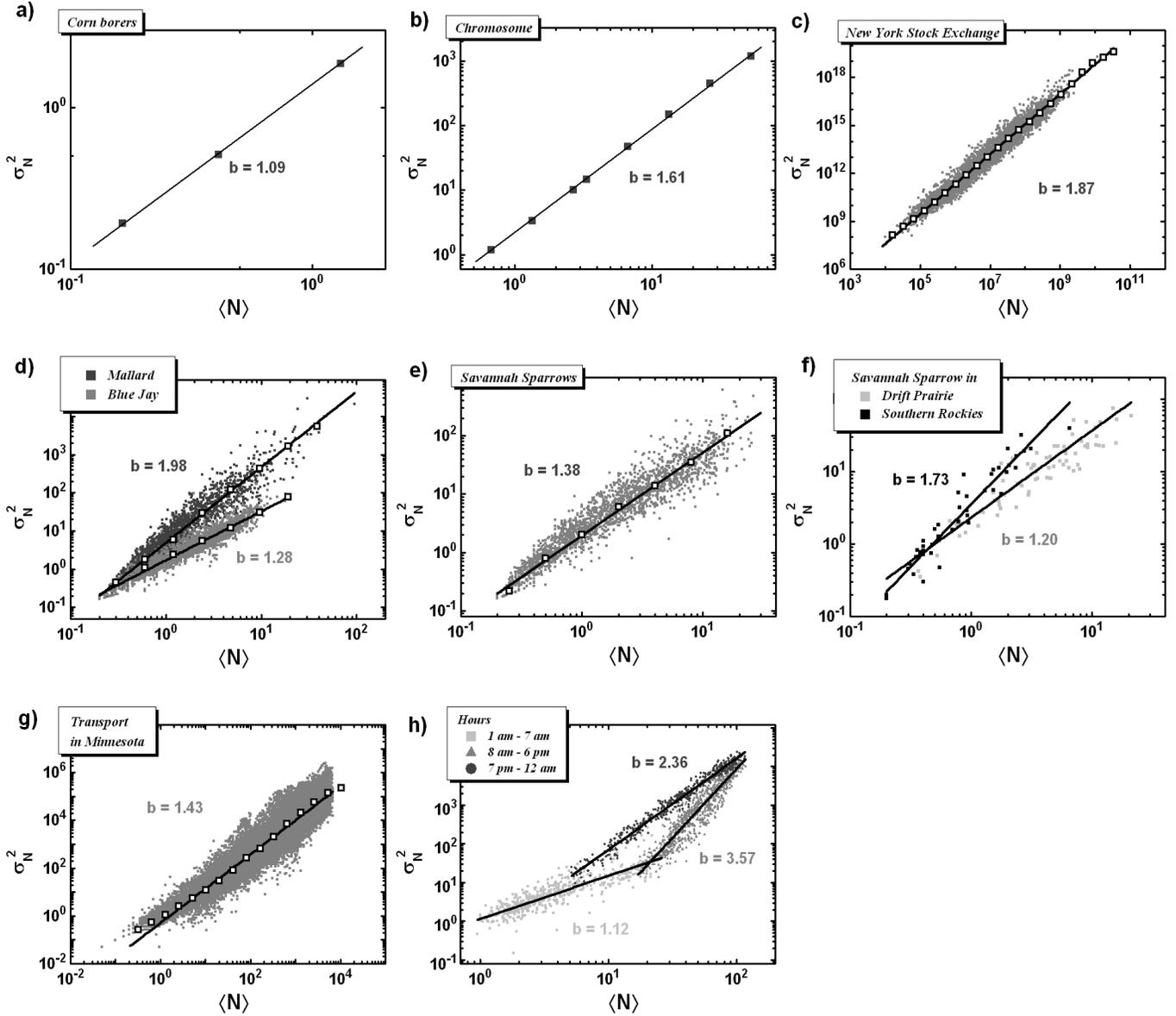}\caption{\label{fig1}{\bf Taylor's fluctuation scaling for systems analyzed in the paper.} {\bf a}, European corn borers (\textit{Pyrausta nubilalis}). {\bf b}, Gene structures in human chromosome 7. {\bf c}, Daily turnovers of stocks traded on the New York Stock Exchange (NYSE). {\bf d}, Mallards ({\it Anas platyrhynchos}) and Blue Jays ({\it Cyanocitta cristata}) in all routes populated by these species in $1966-2007$. {\bf e}, Savannah Sparrows ({\it Passerculus sandwichensis}) in all routes populated by this  species in $1966-2007$. {\bf f}, Savannah Sparrows counted in routes corresponding to two different physiographic conditions located in the Drift Prairie and the Southern Rockies. {\bf g}, Traffic intensity as measured by all automatic traffic recorders in Minnesota in $2007$. {\bf h}, Daily fluctuations in traffic measured by a single recorder (ATR no. $222$) in $2002-2007$. In the figure, full points correspond to raw data (i.e., mean and variance calculated according to the description given in the text and in the Table I), the open symbols express logarithmic binning of the data, and the solid lines represent their linear fits (in log-log scale). In the case of birds, the key observation is that the characteristic parameter $b$ may differ not only among the species but also within the same species if one takes into account the physiographic stratification characterizing living conditions in different areas (for that reason for further analysis we have selected Blue Jays as a species that weakly depends on the physiographic stratification). A similar comment is true for the traffic intensity. In the latter case, one can see that although the whole data obey Taylor's fluctuation scaling with the parameter $b=1.43$, in reality the data are very heterogeneous. The parameter $b$ characterizing traffic intensity recorded by a single ATR may change dramatically from hour to hour.}
\end{figure*}
\squeezetable
\begin{table*}
\begin{tabular}{|p{20mm}|p{9mm}|p{28mm}|p{29mm}|p{76mm}|}
\hline
 \begin{center}Panel in Fig. 1\end{center}&\begin{center}No. of points in the panel\end{center} & \begin{center}No. of data from which each point in the panel (i.e. mean and variance) has been calculated\end{center}&\begin{center}Individual data represents\end{center}&\begin{center}Comments\end{center}\\
\hline
  \begin{center}\vspace{-7pt}1a\end{center} & \begin{center}\vspace{-7pt}\hfill 3\end{center}  & \begin{center}\vspace{-7pt}1296\end{center} & Total number of corn borers living on a given plant& Each point in the panel represents mean and variance for one of three areas each consisted of approximately 3 acre, chosen from the same corn field.\\
\hline
  \begin{center}1b\end{center} & \begin{center}\hfill 8\end{center}  & \begin{center}40 - 3168 (variable)\end{center} & Total number of genes positions of which start in a given bin&Chromosome 7 spans about 159 million DNA building blocks (base pairs) and contains more that 2100 genes (including pseudogenes). The chromosome has been divided into equal-sized non-overlapping bins that spanned the physical length of the chromosome. Number of data from which mean and variance have been calculated depends on the width of the bins (from $5\times 10^4$ to $4\times 10^6$). \\
\hline
  \begin{center}\vspace{-10pt}1c\end{center} & \begin{center}\vspace{-10pt}\hfill 4728\end{center}  & \begin{center}\vspace{-10pt}10\end{center} & Daily turnover of a given stock&Daily turnovers of 2364 stocks were retrieved over two periods of ten consecutive trading days.\\
\hline
\multirow{5}{*}{\hspace{-2pt}\begin{tabular}{p{20mm}}\hspace{8pt}\mbox{1d (Mallard)} \\ \hline \hspace{5pt}\mbox{1d (Blue Jay)} \\ \hline \hspace{25pt}1e \\ \hline \hspace{8pt}\mbox{1f (Prairie)}\\ \hline \hspace{8pt}\mbox{1f (Rockies)} \end{tabular}}& \multirow{5}{*}{\hspace{-2pt}\begin{tabular}{p{9mm}} \hfill 3310 \hfill \\ \hline \hfill 3017\\ \hline \hfill 2428\\ \hline \hfill 73\\ \hline \hfill 51\end{tabular}}& \begin{center}5 - 210 (variable)\end{center}&Total number of individuals of the species recorded in a given part of a given route in a given year&Each route has been divided into 5 parts (stops 1-10, 11-20, 21-30, 31-40, and 41-50). Maximal value in column 3 follows from the calculation: 210 = (5 stops on a route) $\times$ (41 years). Because some species were not recorded each year this quantity can be lower. The value in column 2 is just a number of routes.\\
\hline

\begin{center}1g\end{center} & \begin{center}\hfill 41472 \end{center}& \begin{center}20 - 23 (variable)\end{center} & Total number of cars passing a given ATR in a given hour in one direction in a given weekday of a month of a given year & The value in column 3 is equal to the number of weekdays in a month (so it can vary from 20 days to 23 days depending on the month). The value in column 2 follows from the calculation: 41472 = (72 ATRs) $\times$ (2 directions) $\times$ (24 hours) $\times$ (12 months) $\times$ (1 year)\\
\cline{1-2} \cline{5-5}
\begin{center}\vspace{-10pt}\hspace{6pt}1h (1am-7am)\end{center} & \begin{center}\vspace{-10pt}\hfill 504 \end{center}& & & Value in column 2 follows from the calculation: 504 = (1 ATR) $\times$ (1 direction) $\times$ (7 hours) $\times$ (12 months) $\times$ (6 years)\\
\cline{1-2} \cline{5-5}
\begin{center}\vspace{-10pt}\hspace{5pt}1h (8am-6pm)\end{center} & \begin{center}\vspace{-10pt} \hfill 792 \end{center}&&& Value in column 2 follows from the calculation: 792 = (1 ATR) $\times$ (1 direction) $\times$ (11 hours) $\times$ (12 months) $\times$ (6 years)\\
\cline{1-2} \cline{5-5}
\begin{center}\vspace{-10pt}\hspace{3pt}1h (7pm-12am) \end{center}& \begin{center}\vspace{-10pt}\hfill 432 \end{center} &&& Value in column 2 follows from the calculation: 432 = (1 ATR) $\times$ (1 direction) $\times$ (6 hours) $\times$ (12 months) $\times$ (6 years) \\
\hline

\end{tabular}
\caption{Detailed description of data used to compile Figure 1.}
\end{table*}

Here, we briefly describe the datasets used to validate our theoretical approach. We also draw readers' attention to some known problems (or new observations), which are somehow related to Taylor's law and can be explained in terms of the theoretical framework.

\textit{European corn borer.} A few equilibrium, frequency distributions describing larval populations of this pest (\textit{Pyrausta nubilalis}) have been published in a 1957 paper in Biometrics \cite{1957BiometricsMcGuire}. The paper has brought up an intriguing (and until now unsolved) issue of what kind of frequency distribution should be used to describe different populations. In the study, three areas, each consisted of approximately $3$ acre, chosen from the same corn field, have been investigated. For each area a distribution of larvae per plant has been obtained. In order to fit the gathered data its authors considered three different compound Poisson distributions: the negative binomial, the Neyman type A, and a distribution they called Poisson binomial. Before and a long time after its publication, the distributions described in the paper have been used to characterize different diversity patterns in population biology \cite{2002ApplEcolHolt}. Later in this article, we derive a correct formula for the frequency distribution describing the populations that reveal Taylor's law of the mean.

\textit{Human genome.} Data on physical distribution of gene structures on human chromosome 7
were retrieved from a database provided by the Chromosome 7 Annotation Project \footnote{http://www.chr7.org (The Chromosome 7 Annotation Project).}. Previous
analysis of the data has demonstrated that the density of gene structures within the chromosome
is heterogeneous \cite{2003ScienceScherer}. It has been also shown that the number of such structures contained within a sequence of equal-sized non-overlapping bins that span the physical length of the chromosome fulfills Taylor's fluctuation scaling with the characteristic exponent, $b = 1.61$ \cite{2004Kendal}. This observation has been recognized as a quantitative test confirming the presence of gene clustering within the human genome. In this article, we show that Taylor's law with $b>1$ is always due to collective phenomena such as clustering effects. The case of $b=1$ corresponds to complete randomness described by the Poisson distribution.

\textit{Traffic in transportation networks.} Hourly numbers of cars passing through observation
points located on interstates, trunk highways, county state-aid highways, and municipal state-aid
streets at various locations throughout Minnesota were retrieved from the Minnesota Department
of Transportation \footnote{http://www.dot.state.mn.us/traffic/data/atr/atr.html (The Minnesota Department of Transportation).}. The traffic intensity had been recorded by 72 automatic traffic recorders (ATR) from 2002 to 2007. The datasets, from which mean and variance were calculated, include the number of cars observed by a single recorder and passing in only one direction at a given hour in all weekdays of a month \cite{fronczak2010Acta}. We have checked that such time series have a stationary character (i.e. traffic is homogeneous within the considered periods). In Figure~\ref{fig1}, one can see that although the whole data obey Taylor's fluctuation scaling with the parameter $b = 1.43$, in reality, the data are very heterogeneous and the parameter, $b$, characterizing the data recorded by a single ATR may change dramatically from hour to hour.

\textit{North American avifauna.} The data were retrieved from the North American Breeding Bird
Survey \footnote{http://www.mbr-pwrc.usgs.gov/bbs/bbs.html (The North American Breeding Bird Survey).}. The survey has collected annual abundances over a 40-year period for over 600 bird
species in more than 3000 observation routes. In our analysis, we have concentrated on single
species detected in consecutive years on separate routes. Given such a time series (i.e. the number of birds of a given species counted in a given route in consecutive years), one can calculate its mean and variance (it has been proved elsewhere that the considered time series have a stable temporal structure, see \cite{1998NatureKeitt}). We have done so for three rather abundant species representing different bird families: Mallard (\textit{Anas platyrhynchos}, Family Anatidae), Blue Jay (\textit{Cyanocitta cristata}, Family Corvidae), and Savannah Sparrow (\textit{Passerculus sandwichensis}, Family Emberizidae). Figure~\ref{fig1} presents the empirical Taylor's law for these species. The clue observation is that the characteristic parameter, $b$ (see Eq.~\ref{TaylorLaw}), may differ not only among the species but also within the same species if one takes into account the physiographic stratification characterizing living conditions at different routes (see \cite{1986NatureDowning} for speculations about the meaning and the value of the parameter $b$).

\textit{New York Stock Exchange (NYSE).} Daily turnovers of 2364 stocks were retrieved over two
periods of ten consecutive trading days (May 8 - May 21, 2008 and May 22 - June 5, 2008) from
Yahoo Finance Stock Research Center \footnote{http://biz.yahoo.com/r/ (The Yahoo Finance Stock Research Center).}. The periods were selected due to relative stability of the
NYSE Composite Index and the trading activity in the market. Mean and variance were separately
calculated for each period and each stock. Taylor's scaling exponent of the system, $b = 1.87$, was
quite large. We show that the large value of the parameter, $b$, and Levy flights, which are commonly
observed in stock market dynamics \cite{1995NatureMontegna,BookEcono}, are related to each other, as they result from non-trivial properties of the phase space underlying the considered systems.

Detailed description of how the data were processed is given in Table I and in Appendix~\ref{AppendixData}.

\section{Ensemble fluctuation scaling}

To explain the origins of Taylor's power law for ensemble FS in different complex systems, we start with equilibrium statistical physics. It is well known that the distribution $\mathcal{P}(\Omega;\mu)$ constrained to yield the average value of the parameter $N$ is given by~\cite{BookAttard}
\begin{equation}\label{POmega}
\mathcal{P}(\Omega;\mu)=\frac{e^{-\mu N(\Omega)}}{e^{F(\mu)}},
\end{equation}
where $\mu$ stands for the external field coupled to $N$ that imposes a given value of $\langle N\rangle$ and $F(\mu)$ represents the so-called free energy of the considered system that encodes properties of the system in equilibrium.

\begin{figure}
\includegraphics[width=8.5cm]{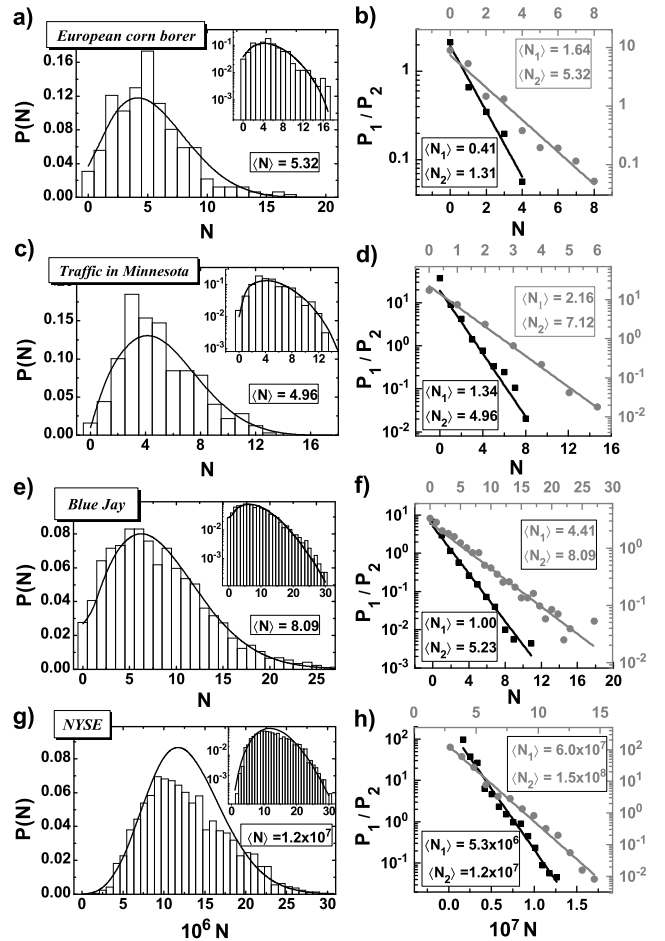}
\caption{\label{fig2}{\bf Comparison of real-world and theoretical frequency distributions $P(N)$ characterizing the considered systems for the given values of $\langle N\rangle$ (left column), and quotients of two distributions fitted by the exponential function predicted by Eq.~(\ref{test1}) (right column).} The figure consists of $8$ panels arranged in $4$ rows. Each row corresponds to a different dataset and provides information on (for detailed information on the data used see Section~\ref{Data} and Appendix~\ref{AppendixData}, detailed description of how fitting of experimental frequency distributions has been done is given in Appendix~\ref{fitting}): {\bf a, b,} European corn borer frequency patterns; {\bf c, d,} Intensity of car traffic in Minnesota, as measured by a single recorder (ATR no. $222$) during night hours ($1$ a.m. - $7$ a.m.) in $2002-2007$; {\bf e, f,} Blue Jay (Cyanocitta cristata) abundance in North America in $1966-2007$; {\bf h, g,} Daily activity at the New York Stock Exchange (NYSE) in the period May $8$ - June $5$, $2008$. Some discrepancies visible in the last case may be due to the problem with selection of more homogeneous subset of analyzed stocks.}
\end{figure}

Here, the crucial point to understand is that the Greek letter $\Omega$ in Eq.~(\ref{POmega}), refers to the so-called microstate of the considered system, whereas researchers studying real-world systems are usually interested in macrostates and the corresponding macroscopic quantities. Accordingly, in this investigation, instead of the distribution $\mathcal{P}(\Omega;\mu)$, we seek the expression for the frequency distribution $P(N;\mu)$ (sometimes we write $P(N)$ instead of $P(N;\mu)$) that characterizes macroscopic states of different systems fulfilling Taylor's law. The essence of our approach lies in a rather trivial observation that the two introduced distributions are related to each other by a simple expression,
\begin{equation}\label{defDOS}
P(N;\mu)=g(N)\mathcal{P}(\Omega;\mu),
\end{equation}
where $g(N)$ is the announced NoS, which gives the number of microstates $\Omega$ having the same value of the macroscopic parameter $N$. In the following, we show that in the systems fulfilling Taylor's FS, the function $g(N)$ has a well-defined form, and this immediately allows us to infer the origin of Taylor's law to understand the meaning of Taylor's characteristic parameter, $b$.

Provided that $N$ is nonnegative and discrete, the normalizing factor in Eq.~(\ref{POmega}), $e^{F(\mu)}$, is just the Z-transform of NoS, $g(N)$,
\begin{equation}\label{SdefF}
e^{F(\mu)}=\sum_\Omega e^{-\mu N(\Omega)}=\sum_{N=0}^{\infty}g(N)e^{-\mu N}.
\end{equation}
Using the exponential formula known from combinatorial mathematics \cite{BookStanley}, the left-hand side of Eq.~(\ref{SdefF}) can be written as
\begin{equation}\label{Sstep1}
e^{F(\mu)}= 2e^{f_0}+\sum_{N=1}^{\infty}\left[\frac{e^{f_0}}{N!}B_N(f_1,f_2,\dots,f_N)\right]\;\mu^N,
\end{equation}
where $B_N(f_1, ..., f_N)$ is the $N$th complete Bell polynomial whereas $f_n$ represents the coefficient of the $n$th term in the MacLaurin expansion of the free energy,
\begin{equation}\label{Fenergy}
F(\mu)=\sum_{n=0}^\infty \frac{f_n}{n!}\mu^n.
\end{equation}
In accordance with Eq.~(\ref{Sstep1}), the right-hand side of Eq.~(\ref{SdefF}) can be written in the following form
\begin{equation}\label{Sstep2}
\sum_{N=0}^\infty g(N)e^{-\mu N}=\sum_{n=0}^\infty (-1)^n\left[\sum_{N=0}^\infty \frac{N^n}{n!}g(N)\right]\;\mu^n.
\end{equation}
Although the expressions in (\ref{Sstep1}) and (\ref{Sstep2}) seem to be rather complicated, their theoretical and real-world interpretation is very simple.

Let us concentrate on mathematical issues. First, the formula within the square brackets in Eq.~(\ref{Sstep2}) is the Poisson transform,
\begin{equation}\label{new1}
\mathcal{G}(n)=\mbox{P.T.}\left[ G(N),n\right]=\sum_{N=0}^\infty G(N)e^{-N}N^n/n!,
\end{equation}
of the function
\begin{equation}\label{new2}
G(N)=e^Ng(N).
\end{equation}
When working with the Poisson transform, it is important to understand how the transform acts on an arbitrary function. Simplifying, one could say that the transform adds Poissonian fluctuations to the function. For that reason, $\mathcal{G}(n)$ looks like a fuzzy image of the original $G(N)$, and it is often reasonable to assume that (for more details see Appendix~\ref{AppendixPT})
\begin{equation}\label{new3}
\mathcal{G}(n)=\mbox{P.T.}\left[G(N),n\right]\simeq G(n).
\end{equation}

The last property of the transform turns out to be very useful in the inverse problems, in which one has to calculate the original function provided that its Poisson transform is known. Such a problem arises in our derivations. Comparing Eqs.~(\ref{Sstep1}) and (\ref{Sstep2}) for $n\geq 1$ one gets
\begin{equation}\label{SPTDOS2}
\mathcal{G}(n)=e^{f_0}(-1)^n\frac{1}{n!}B_n(f_1,f_2,\dots,f_n).
\end{equation}
The closed form of the inverse Poisson transform for $\mathcal{G}(n)$ and the closed form of $g(N)$ only exist in a few cases. In the general case, for $N\gg 1$, one can use the following approximation
\begin{equation}\label{DOSfinal}
g(N)\simeq e^{-N}\mathcal{G}(N)\propto \frac{e^{-N}}{N!}B_N(f_1,f_2,\dots,f_N).
\end{equation}
Eq.~(\ref{DOSfinal}) is the main theoretical result of this article. It has a lot of in common with the famous Mayer's diagrammatic expansions for imperfect gas \cite{BookHansen,BookAttard}. Further in this paper we show that the derived NoS function has a very intuitive form.

To apply the derived formula to systems with ensemble FS one has to know all the parameters $f_1,f_2,\dots,f_N$, i.e. one has to find $F(\mu)$ describing systems obeying Taylor's law. In order to do it we exploit fluctuation-dissipation relation
\begin{equation}\label{SdefFDR}
\langle N^2\rangle-\langle N\rangle^2=-\frac{\partial \langle N\rangle}{\partial\mu}=\frac{\partial^2F(\mu)}{\partial\mu^2},
\end{equation}
which states that fluctuations of the parameter $N$ are proportional to susceptibility of the parameter to its conjugate field $\mu$. Comparing right-hand sides of Eqs.~(\ref{SdefFDR}) and (\ref{TaylorLaw}) one obtains differential equation for $\langle N\rangle$, i.e., $-\partial \langle N\rangle/\partial\mu= a\langle N\rangle^b$. Solving this equation, with the reasonable assumption of nonnegative variance, one gets
\begin{equation}\label{meanN}
\langle N\rangle=\left\{ \begin{array}{lcl}
Xe^{-a\mu} & \mbox{for } & b=1 \\
((b-1)a\mu+X)^{1/(1-b)} & \mbox{for } & b>1.
\end{array}\right.
\end{equation}
Next, having (\ref{meanN}) and again exploiting (\ref{SdefFDR}), i.e., solving $\partial F/\partial\mu=-\langle N\rangle$, one also finds the formula for the free energy, $F(\mu)$,
\begin{equation}\label{FreeEnergy}
F(\mu)=\left\{ \begin{array}{lcl}
\frac{1}{a(2- b)}\langle N\rangle^{(2-b)}+Y & \mbox{for } & b\geq 1 \\\frac{1}{a}\ln\langle N\rangle+Y & \mbox{for } & b=2.
\end{array}\right.
\end{equation}
$X$ and $Y$ represent integration constants in Eqs. (\ref{meanN}) and (\ref{FreeEnergy}). Coefficients in the MacLaurin  expansion (\ref{Fenergy}) of the free energy $F(\mu)$ are given by (for derivation see Appendix~\ref{Appendixfn})
\begin{equation}\label{fn}
f_n=x_n\langle N\rangle^{(n-1)b-(n-2)}|_{\mu=0},
\end{equation}
where
\begin{equation}\label{fnx}
x_n=(-1)^na^{n-1}\prod_{i=2}^n[(n-2)b-(n-3)].
\end{equation}
Detailed calculations for the Poisson distribution $P(N;\mu)$ (\ref{defDOS}) are shown in Appendix \ref{AppendixPoisson}.

Now, before delving into an interpretation of the derived expressions, it would be valuable to convince the reader that our approach does really account for the behavior of real-world systems with ensemble FS. To do so, we performed three quantitative tests on experimental data. The tests clearly show that Taylor's law results from the unique NoS underlying the considered systems.

\begin{figure}
\includegraphics[width=8cm]{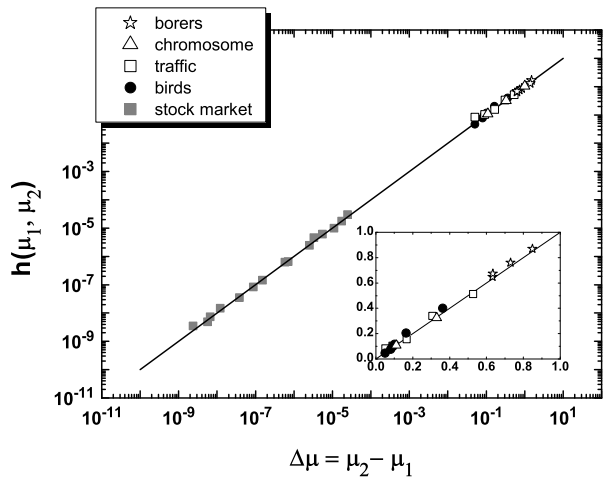}
\caption{\label{fig3}{\bf Experimental verification of Eq.~(\ref{test2}).} Different symbols placed in the graph represent different real-world systems that show Taylor's fluctuation scaling. Note that the symbols cover the solid theoretical line to an impressive extent of nine orders of magnitude of $\Delta\mu$.}
\end{figure}

The first test consists of a direct comparison between experimental frequency distributions and theoretical distributions given by Eq.~(\ref{defDOS}). The second test shows that the quotient of two frequency distributions corresponding to different average values of $\langle N\rangle$ is an exponential function of $N$,
\begin{equation}\label{test1}
\frac{P(N;\mu_1)}{P(N;\mu_2)}\propto e^{(\mu_2-\mu_1)N}.
\end{equation}
The third test follows from Eq.~(\ref{meanN}) and indirectly refers to Eq.~(\ref{test1}). Namely, transforming the expression for $\langle N\rangle$, one can show that the experimental data should satisfy the identity
\begin{equation}\label{test2}
\mu_2-\mu_1=\frac{\langle N_2\rangle^{(1-b)}-\langle N_1\rangle^{(1-b)}}{a(1-b)}=h(\langle N_1\rangle,\langle N_2\rangle).
\end{equation}
Figures \ref{fig2} and  \ref{fig3} present results of the three tests applied to systems with ensemble FS: European corn borers and human genome. Note the excellent agreement between the data and our theoretical predictions.

\section{Temporal fluctuation scaling}

Now, the question is: Does the explanation of origin of ensemble FS in equilibrium systems may help to understand temporal FS characterizing systems in non-equilibrium steady states? The answer is affirmative. Recently, it was shown that phase space probability distribution describing such systems has an exponential form \cite{Attard5,AttardRev}, cf.~Eq.~(\ref{POmega}),
\begin{equation}\label{attard1}
\mathcal{P}(\Omega;\mu,\mu_1)=\frac{e^{-\mu N(\Omega)}\;e^{-\mu_1(N_1(\Omega)+W_1^{mir})}}{e^{F(\mu,\mu_1)}},
\end{equation}
where  $N_1$ is the so-called first moment of an additive variable $N$ that the considered system may exchange with the reservoir. The parameters $\mu$ and $\mu_1$ stand for external fields coupled to $N$ and $N_1$ respectively, whereas $W_1^{mir}$ is called the mirror work. Finally, $F(\mu,\mu_1)$ represents nonequilibrium free energy (see also \cite{2006Sasa}).

The formula (\ref{attard1}) derived by Phil Attard generalizes the Boltzmann distribution, Eq.~(\ref{POmega}), to nonequilibrium systems. To account for temporal FS described by Eq.~(\ref{TaylorLaw}), we use the formula to calculate frequency distribution $P(N;\mu,\mu_1)$. Similarly as in the case of equilibrium statistical physics, to get $P(N;\mu,\mu_1)$ one has to sum Eq.~(\ref{attard1}) over all microstates, $\Omega^*$, that fulfill the condition $N(\Omega^*)=const$, i.e.
\begin{equation}\label{apf1}
P(N;\mu,\mu_1)=\sum_{\Omega^*}\mathcal{P}(\Omega;\mu,\mu_1)=\frac{e^{-\mu N}}{e^{F(\mu,\mu_1)}}g^*(N;\mu_1),
\end{equation}
where $g^*(N;\mu_1)=\sum_{\Omega^*}e^{-\mu_1(N_1+W_1^{mir})}$ may be termed as the weighted NoS. Finally, averaging Eq.~(\ref{apf1}) over different values of the parameter $\mu_1$ one can reduce it to the formula that is equivalent to Eq.~(\ref{defDOS})
\begin{eqnarray}\label{apf2}
P(N;\mu)&=&\langle P(N;\mu,\mu_1)\rangle=P(N;\mu,\mu_1^*)\\ \nonumber &=&\frac{e^{-\mu N}}{e^{F(\mu,\mu_1^*)}}g^*(N;\mu_1^*),
\end{eqnarray}
where the last transformation in (\ref{apf2}) is valid under the generalized mean value theorem and $\mu_1^*$ is a certain value of the parameter $\mu_1$.

The reasoning behind the algebra is that the formalism applied to ensemble FS may be adopted to explain temporal fluctuations. Results of three tests (as described earlier) applied to systems with temporal FS (i.e. traffic intensity in Minnesota, dynamics of North American breeding bird populations and NYSE) shown in Figs.~\ref{fig2} and \ref{fig3} certify the statement.

\section{Number of states in real-world systems}

In the following, we concentrate on the formula for the number of states, $g(N)$ (\ref{DOSfinal}). Although the formula seems rather complicated, its theoretical and real-world interpretation is very simple. In general, the $N$th complete Bell polynomial, $B_N(f_1,f_2,\dots,f_N)$, describes the number of disjoint partitions of a set of size $N$ into an arbitrary number of subsets~\cite{Bellinfo1,Bellinfo2}. The parameters $f_i$ with $i=1,2,\dots,N$ apply to subsets of size $i$ and play an important role in a description of the partitions. For example, if all the parameters $f_1,f_2,\dots,f_N$ have the same value, then there is no preference for the size of subsets. The resulting partitions correspond to a random distribution of elements, and the frequency distribution describing that system, $P(N)$ (\ref{defDOS}), is Poissonian (for a detailed derivation see Appendix~\ref{AppendixPoisson}). On the other hand, in the extreme case of $f_{i}\gg f_{j}$ for all $j\neq i$, the Bell polynomial gives the number of such partitions in which there is a strong preference for subsets of size $i$.

\begin{figure*}
\includegraphics[width=0.9\textwidth]{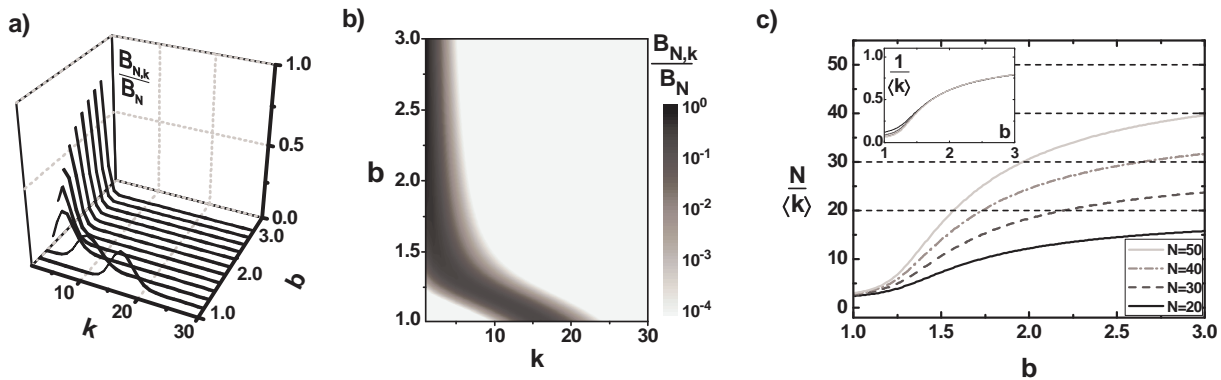}
\caption{\label{fig4}{\bf Interpretation of the number of states function $g(N)$ given by Eq.~(\ref{DOSfinal}).} Decomposition of the complete Bell polynomials into partial Bell polynomials for different values of Taylor's parameter $b$. We have already shown that different values of $b$ result from different free energies $F(\mu)$ (\ref{FreeEnergy}) characterizing the considered systems, which, in turn, provide different sets of the coefficients $f_1,f_2,\dots,f_N$. Graphs {\bf a} and {\bf b} show that the whole set of partitions for a given $b$ is dominated by partitions that consist of a well-defined number of subsets, $k$ (here, $N=50$). This number decreases with $b$ leading to an increase in the average size of these subsets, $N/\langle k\rangle$ (cf. graph {\bf c}). The decomposition analysis clearly shows that higher values of $b$ correspond to aggregation effects. The elements of the original set of size $N$ aggregate into subsets whose sizes increase with $b$.}
\end{figure*}

Given the meaning of the complete Bell polynomials, an interpretation of the number of states (\ref{DOSfinal}) underlying real-world systems with Taylor's fluctuation scaling follows immediately. For example, in the case of the bird population of size $N$, the number of states, $g(N)$, is proportional to the number of different partitions of $N$ birds into subpopulations of arbitrary size. The number and the size of subpopulations are encoded in both the free energy of the system, $F(\mu)$, and the corresponding parameters, $f_i$. The analogous interpretation of the NoS applies to every other animal, insect, and plant population.

In the case of stock market dynamics, traffic in transportation networks, and other systems driven by human activity, the number of states in (\ref{DOSfinal}) has a similar interpretation. In order to show the analogy, let us concentrate on the number of cars (e.g., $N=60$) passing through a given observation point in a given time period (e.g., $T=60$ minutes). First, note that the parameters $N$ and $T$ do not tell the whole story about the traffic. The same values of $N$ and $T$ may result from homogeneous traffic (e.g., on average one car per one minute) or inhomogeneous traffic (e.g., all cars counted within the first five minutes). According to our approach, all microstates $\Omega$ with the same $N$ are equiprobable as shown in (\ref{POmega}). This does not mean, however, that all the states are possible from the point of view of the considered system. For example, if the observation point is located on a very busy roadway, then states of the roadway with no cars running are unlikely.

In fact, the problem of car traffic may be simplified to the problem of $N$ balls in $T$ boxes (boxes may represent minutes with balls corresponding to cars). In this notation, the meaning of the number of states given by the complete Bell polynomials is clear. The number of states, $g(N)$, corresponds to the number of different partitions of $N$ balls (cars) into $T$ boxes (minutes). The parameters $f_1,f_2,\dots,f_N$ indicate which partitions are reasonable and likely from the point of view of the considered system. If the balls are noninteracting, all the parameters $f_i$ have the same value. On the other hand, when the spectrum of $f_i$ is not uniform, there must exist interaction between the balls (cars) resulting in different collective phenomena, e.g., clustering seen as a traffic jam.

To complete the discussion of the number of states given by Eq.~(\ref{DOSfinal}), one should mention the so-called partial Bell polynomials, $B_{N,k}(f_1,f_2,\dots,f_{N-k+1})$~\cite{Bellinfo2}. The polynomials describe the number of partitions of a set of size $N$ in which exactly $k$ subsets are considered, $B_N(f_1,f_2,\dots,f_N)=\sum_{k=1}^N B_{N,k}(f_1,f_2,\dots,f_{N-k+1})$. The decomposition of the complete Bell polynomials into partial polynomials allows one to analyse how different partitions contribute to the number of states for different values of the parameter $b$.

In Figure~\ref{fig4}, one can see such a decomposition for different values of $N$ and $1\leq b\leq 3$. It is remarkable that the preferred number of subsets, $\langle k\rangle$, and the preferred size of these subsets, $N/\langle k\rangle$, have well-defined values, and these values strongly depend on the parameter $b$ of Taylor's power law (\ref{TaylorLaw}). The average number of subsets is a decreasing function of  $b$ and, similarly, the average size of subsets is an increasing function of $b$. This means that larger values of the parameter characterise stronger collective phenomena manifesting either in aggregations of individuals in different populations~\cite{1961NatureTaylor,1986NatureDowning}, traffic jams~\cite{2001RMPHelbing}, and Levy flights in stock market dynamics~\cite{BookEcono}. This supports the scientific message of this article; that Taylor's fluctuation scaling is due to the number of states underlying the considered systems. The number of states has a built-in susceptibility of the system to collective effects, the strength of which depends on the value of Taylor's parameter, $b$. It also means that one need not invoke any stochastic models to explain these phenomena. In fact, there may exists a number of stochastic processes defined in the phase space with NoS given by Eqs.~(\ref{DOSfinal}) and (\ref{FreeEnergy}). Of course, all the processes will result in Taylor's fluctuation scaling.

\section{Concluding remarks}

In closing, we note that our approach differs crucially from previous work on Taylor's power law~\cite{1977NatureTaylor, 1982NatureAnderson,1983NatureTaylor,1986NatureDowning,1986GJTBGills,1994ProcPerry,
2003NatureKilpartick,2004EcolComplexKendal,2004PRLMenezes2}. It is significant that the approach does not invalidate the previous models. On the contrary, our approach may allow identification of the models' important features that account for the law. We show that Taylor's fluctuation scaling results from the ubiquitous second law of thermodynamics (here called the maximum entropy principle) and the number of states (a concept borrowed from physics). We anticipate that our formalism will provide a quantitative basis for formulating a new theory of populations, communities, and ecosystems~\cite{2002PTBSBrown}, a theory based on concepts of the number of states. In epidemiology and medicine, our explanation of Taylor's law may be helpful in accounting for such observations as clustering in human sexual contact in HIV transmission~\cite{1989NatureAnderson}, epidemic outbreaks~\cite{1996NatureRhodes}, or high variability of cancer statistics within the human population~\cite{2007MathBiolKendal}. We believe that the approach may also help in understanding the variability in organ cell numbers for a variety of organisms~\cite{2001PNASAzevedo} and the clustering of genes on human chromosomes~\cite{2004Kendal}. In all the listed cases, Taylor's law has been recognized as an intrinsic feature of the considered systems, and the new interpretation of the law offered here may help clarify the underlying dynamics.

\begin{acknowledgments}
We acknowledge financial support from the Ministry of Education and Science in Poland, under grants ESF/275/2006 (AF) and 496/N-COST/2009/0 (PF). We also thank Prof. Z. Burda for helpful discussions.
\end{acknowledgments}

\appendix

\section{Data processing}\label{AppendixData}

We would like to stress that although there are plenty of real data in which Taylor's fluctuation scaling can be observed, most of the data require careful processing before they can be used in our analysis. The two basic problems that we encountered in trying to validate our theoretical approach in real-world systems were related to: heterogeneity of the considered data and/or small amount of the data.

The first problem is partially illustrated in Fig.~\ref{fig1}. The essence of the problem is that, although the bulk of the considered data fulfills Taylor's law with a given parameter b, in fact, the
data may consist of a certain number of subsets $i = 1,2,\dots$ (e.g., representing logical subsystems
of the considered system), each of which is characterized by its own characteristic parameter, $b_i$.
Careless selection of data may lead to misleading results. The quantitative tests basing on Eqs.~(\ref{defDOS}), (\ref{test1}) and (\ref{test2}) described in the article are peculiarly sensitive to such a careless attitude to data. Therefore, in most cases, data standardization is required. Unfortunately, such a standardization may significantly decrease the amount of data. It may even lead to questions on statistical significance of the preformed analysis.

The problem of statistical significance is the more important because, although Taylor's law
(\ref{TaylorLaw}) operates on macroscopic parameters (i.e., on the mean, $\langle N\rangle$, and the variance, $\sigma^2_N=\langle N^2\rangle-\langle N\rangle^2$, of the frequency distribution, $P(N)$), from both experimental and theoretical points of view the basic observable in our approach is the distribution, $P(N)$, itself. To prove validity of our approach, one has to operate on large datasets, which provide smooth experimental frequency distributions, $P(N)$, and are also homogeneous in the sense of the parameter, $b$.

For these reasons, the actual verification of our theoretical approach shown in Figs.~\ref{fig2} and \ref{fig3} in the main body of the article was, inter alia, based on the most abundant bird species representing the North American avifauna and on the frequency distributions describing larval populations of European corn borer published in a classical paper \cite{1957BiometricsMcGuire}. In analyzing traffic intensity in Minnesota, we have concentrated on only one (from among 72 others) traffic recorder in night hours. To increase the amount of data, we have analyzed traffic over a very long time period of 5 years (2002 - 2007). Analysis of such a long time period was possible due to relative stability of traffic measured by the considered recorder during this time period. A similar procedure of data selection has been applied to the NYSE data. However, although we expected the system to be very heterogeneous in the sense of Taylor's parameter, $b$, we did not observe any of these effects. Contrary to transportation network, our only concern with the stock market was the highly nonstationary character of the data. Therefore, a more detailed analysis of NYSE was based on daily turnovers
of all 2364 stocks quoted in this market in a rather short time period 2$\times$10 consecutive trading
days (i.e. May 8 - May 21, 2008 and May 22 June 5, 2008). The periods correspond to relative
stability on both trading activity in the market and the NYSE Composite Index 39.

\begin{figure}
\includegraphics[width=7cm]{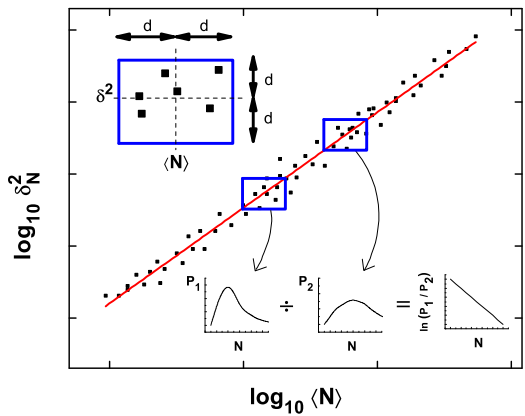}
\caption{\label{figS1} {\bf (Color online). Schematic description of the methodology used in real data analysis.} Detailed description is given in the text.}
\end{figure}

The above description of data selection together with the general description of the considered
datasets should make possible independent reconstruction of Fig.~\ref{fig1}. Below, we describe our
method to obtain the smooth experimental frequency distributions, $P(N)$, shown in Fig.~\ref{fig2}.

We were interested in smooth frequency distributions, $P(N)$, possessing the given mean,
$\langle N\rangle$, and the corresponding variance, $\sigma^2_N=\langle N^2\rangle-\langle N\rangle^2$. Having the mean-variance graphs, as shown in Fig.~\ref{fig1}, the smooth distributions characterizing single points in these graphs can be only observed in two cases of distributions describing larval populations of European corn borer and gene structures in the human chromosome 7. In the remaining cases (including bird populations, NYSE, and transportation network), distributions corresponding to single points on the mean-variance graph are very noisy. To increase the amount of data from which a distribution is made, we have assumed that the neighboring points in the mean-variance graph result from similar environmental conditions (i.e., from similar values of the parameter $\mu$). In this way, a single smooth distribution possessing the given values of $\langle N\rangle$ and $\sigma^2_N$ has been prepared as a simple sum of all the component distributions corresponding to single points in the mean-variance graph and meeting the following conditions:
\begin{equation}
\left|\log_{10}\frac{\langle N_i\rangle}{\langle N\rangle}\right|<\log_{10} d
\end{equation}
and
\begin{equation}
\left|\log_{10}\frac{\sigma^2_{N,i}}{\sigma^2_N}\right|<\log_{10} d
\end{equation}

\begin{figure}
\includegraphics{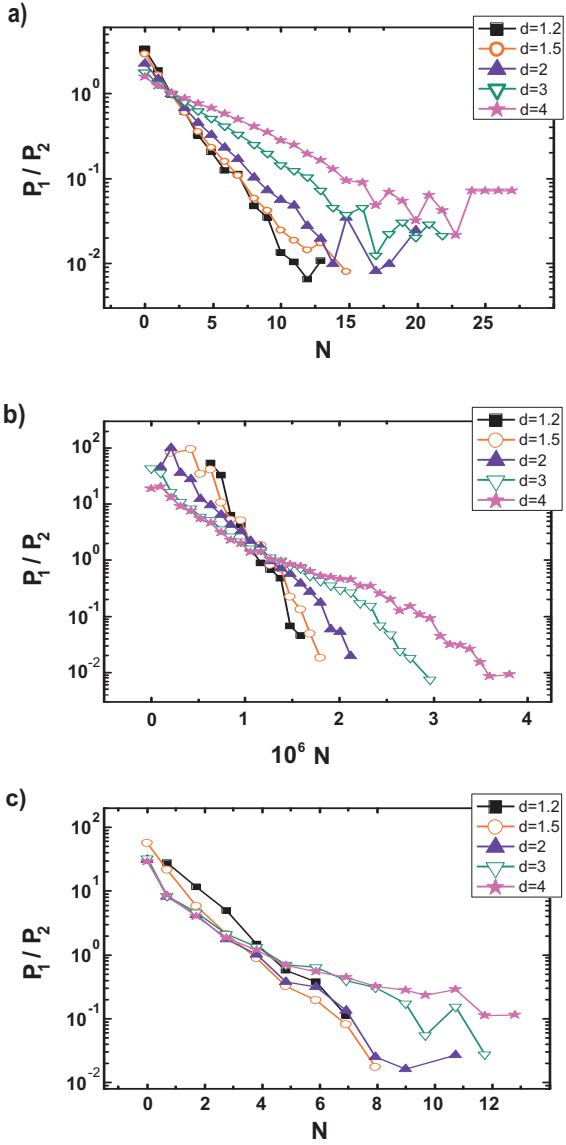}
\caption{\label{figS2} {\bf (Color online). Dependence of the quotients, $P_1(N)/P_2(N)$, on the parameter $d$.} {\bf a,} Blue Jay ({\it Cyanocitta cristata}) populations in North America, $\langle N_1\rangle=1.22$ and $\langle N_2\rangle=3.71$. {\bf b,} Dynamics of the NYSE, $\langle N_1\rangle=6.62\times 10^5$ and $\langle N_2\rangle=2.26\times 10^6$. {\bf c,} Traffic intensity in Minnesota, $\langle N_1\rangle=2.16$ and $\langle N_2\rangle=8.59$.}
\end{figure}

Interpretation of the parameter $d$ is easy. It describes a linear size of the square in the log-log
plot in the mean-variance graph with the central point of the square, $[\langle N\rangle,\sigma^2_N]$, placed in the solid line corresponding to the empirical Taylor's law (cf. Fig.~\ref{figS1}). If $d$ is chosen to be too small, i.e., $d\rightarrow 1$, then the square shrinks to the single point, $[\langle N\rangle,\sigma^2_N]$. It leads to poor data sampling and noisy distribution, $P(N)$. On the other hand, when the parameter $d$ is taken too large, the resulting distribution, $P(N)$, is made of distributions that characterize rather different environmental conditions. Although the obtained distribution is smooth, it is not very reliable. Therefore, a proper choice of the parameter $d$ is important. For the three datasets considered in this article (i.e., North American avifauna, NYSE, and transportation network in Minnesota) we have chosen $d=1.5$. We have checked that, in all the considered cases, the value is optimal. With $d=1.5$, the shape of frequency distributions is already sufficiently smooth and the obtained quotients, $P_1(N)/P_2(N)$, do not change significantly compared with the smaller values of d (see Fig.~\ref{figS2}).

For better understanding of data processing we have developed a web page where the data sets as well as a software for their analysis have been placed~\cite{stronka}.

\section{Supplementary equations}

\subsection{Approximating function by its Poisson Transform}\label{AppendixPT}

Let us consider the Poisson transform defined as follows:
\begin{equation}\label{B0a}
\mathcal{G}(n)=\mbox{P.T.}\left[G(N),n\right]=\sum_{N=0}^\infty G(N)\frac{e^{-N}N^n}{n!}.
\end{equation}
In this appendix, we show that the accuracy of the approximation given by Eq.~(\ref{new3}), i.e.
\begin{equation}\label{B0b}
\mathcal{G}(n)\simeq G(n),
\end{equation}
may be evaluated on the basis the following theorem:

{\bf Approximation theorem for the Poisson transform, as defined by Eq.~(\ref{B0a}) .} {\it Let $\mathcal{G}(n)$ be the Poisson transform of $G(N)$. Then we have
\begin{equation}\label{B1}
G(N)=\mathcal{G}(N)+\sum_{j\geq 1}\sum_{i=j+1}^{2j}c_{i,j}\;\mathcal{G}^{(i)}(N),
\end{equation}
where $\mathcal{G}^{(i)}(N)$ corresponds to the $i$-th derivative of $\mathcal{G}(n)$ at $n=N$ and
\begin{equation}\label{B2}
c_{i,j}=\frac{1}{i!}\sum_{k\geq 0}(-1)^{i-k+j}{i\choose k}\left[{k \atop k-j}\right],
\end{equation}
where $\left[{k \atop k-j}\right]$ stands for Stirling numbers of the first kind.}

One can show that the coefficients $c_{i,j}$ satisfy the following recurrence relation \cite{Poblete1986}
\begin{equation}\label{B3}
(i+1)c_{i+1,j+1}=-ic_{i,j}-c_{i-1,j},
\end{equation}
with boundary conditions $c_{0,j}=\delta_{0,j}$, where $\delta_{i,j}$ stands for the Kronecker's delta, and $c_{1,j}=0$. Table~\ref{tab1} lists some of these coefficients $c_{i,j}$.

\begin{table*}
\begin{center}
\begin{tabular}{cccccccc}
\hline
$_i\backslash\!^j$&&0&1&2&3&4&5\\
\hline
0  &&  1  &  0  &  0  &  0  &  0  &  0\\
1  &&  0  &  0  &  0  &  0  &  0  &  0\\
2  &&  0  &  $-1/2$  &  0  &  0  &  0  &  0\\
3  &&  0  &  0  &  $1/3$  &  0  &  0  &  0\\
4  &&  0  &  0  &  $1/8$  &  $-1/4$  &  0  &  0\\
5  &&  0  &  0  &  0  &  $-1/6$  &  $1/5$  &  0 \\
6  &&  0  &  0  &  0  &  $-1/48$  &  $13/72$  &  $-1/6$\\
7  &&  0  &  0  &  0  &  0  &  $1/24$  &  $-11/60$\\
8  &&  0  &  0  &  0  &  0  &  $1/384$  &  $-17/288$\\
9  &&  0  &  0  &  0  &  0  & 0  &  $-1/144$\\
10  &&  0  &  0  &  0  &  0  &  0  &  $-1/3840$\\
\hline
\end{tabular}
\end{center}
\caption{Values of the coefficients $c_{ij}$.} \label{tab1}
\end{table*}

To prove the theorem we will use an analogous approximation theorem for the Poisson transform defined in the following manner
\begin{equation}\label{B4}
\mathcal{F}_m(n)=\mbox{P.T.}^*\left[F_m(N),n\right]=\sum_{N=0}^\infty F_m(N)\frac{e^{-mn}(mn)^N}{N!}.
\end{equation}
(Note that the Poisson transforms $\mathcal{G}(n)$ (\ref{B0a}) and $\mathcal{F}_m(n)$ (\ref{B4}) differ between each other. The former naturally emerges in many physical  problems, e.g. in optics \cite{Wolf1964} and the science of complex networks \cite{Boguna2003,Fronczak2006}, whereas the latter is widely used in computer science and information theory.) The theorem states \cite{Poblete1986}:

{\bf Approximation theorem for the Poisson transform, as defined by Eq.~(\ref{B4}) .} {\it Let $\mathcal{F}_m(n)$ be the Poisson transform of $F_m(N)$. Then, for $n=N/m$, we have
\begin{equation}\label{B5}
F_m(N)=\mathcal{F}_m(n)+\sum_{j\geq 1}\left(\frac{1}{N}\right)^j\sum_{i=j+1}^{2j}c_{i,j}\;n^i\;\mathcal{F}_m^{(i)}(n),
\end{equation}
where $\mathcal{F}_m^{(i)}(n)$ corresponds to the $i$-th derivative of $\mathcal{F}_m(n)$
and the coefficients $c_{i,j}$ are given by Eq.~(\ref{B2}).}

To prove the approximation theorem for the Poisson transform given by Eq.~(\ref{B0a}) we first put $m=1$ in Eq.~(\ref{B5}). Next, we note that the two transforms, $\mathcal{G}_(n)$ (\ref{B0a}) and $\mathcal{F}_1(n)$ (\ref{B4}) represent the same function, i.e.
\begin{equation}\label{B6}
\mathcal{F}_1(n)=\mathcal{G}(n),
\end{equation}
when
\begin{equation}\label{B7}
F_1(N)=G(N)\;\left(\frac{n^Ne^{-n}}{N!}\right)^{-1}\;\left(\frac{N^ne^{-N}}{n!}\right).
\end{equation}
Then, replacing $F_1(N)$ in Eq.~(\ref{B5}) with Eq.~(\ref{B7}), and using Eq.~(\ref{B6}) we immediately get Eq.~(\ref{B1}). This finishes the proof.

Summarizing, approximating function $G(n)$ by its Poisson transform $\mathcal{G}(n)$, cf. Eq.~(\ref{B0b}), is acceptable when the transform varies slowly enough.

\subsection{Analytical formula for the coefficients $f_n$}\label{Appendixfn}

In the main text of our paper, the parameters, $f_1,f_2,\dots,f_N$ stand for coefficients of the consecutive terms in the MacLaurin expansion (\ref{Fenergy}) of the free energy, $F(\mu)$ (\ref{FreeEnergy}). It means that the single coefficient, $f_n$, corresponds to the $n$th derivative of $F(\mu)$ at $\mu=0$
\begin{equation}\label{afS1}
f_n=F^{(n)}(0).
\end{equation}
Given Taylor's fluctuation scaling (\ref{TaylorLaw}) and the fluctuation-dissipation relation (\ref{SdefFDR}), it is not difficult to derive the general formula for $f_n$. The first steps of this derivation are given below.

First, one has
\begin{equation}\label{afS2}
f_0=F(0).
\end{equation}
The coefficient, $f_1$, simply results from the fluctuation-dissipation relation
\begin{equation}\label{afS3}
f_1=\left.\frac{\partial F(\mu)}{\partial\mu}\right|_{\mu=0}=-\langle N\rangle|_{\mu=0}.
\end{equation}
Having the expression for $f_1$,the coefficient, $f_2$, can be calculated in the following manner
\begin{eqnarray}
f_2&=&\left.\frac{\partial^2 F(\mu)}{\partial\mu^2}\right|_{\mu=0}=\left.\frac{\partial}{\partial\mu} \left(\frac{\partial F(\mu)}{\partial\mu}\right)\right|_{\mu=0}\label{afS4}\\
&=&(\langle N^2\rangle-\langle N\rangle^2)|_{\mu=0}=(a\langle N\rangle^b)|_{\mu=0}\nonumber.
\end{eqnarray}
The next coefficients $f_n$ can be derived in a similar way. In particular, using Eq.~(\ref{afS4}) one gets
\begin{eqnarray}
f_3&=&\left.\frac{\partial^3 F(\mu)}{\partial\mu^3}\right|_{\mu=0}=\left.\frac{\partial}{\partial\mu} \left(\frac{\partial^2 F(\mu)}{\partial\mu^2}\right)\right|_{\mu=0}\label{afS5}\\
&=&\frac{\partial(a\langle N\rangle^b)}{\partial\mu}|_{\mu=0}=\dots=(-1)^2a^2b\langle N\rangle^{(2b-1)}|_{\mu=0}\nonumber.
\end{eqnarray}
Continuing these calculations, one can show that the general formula describing $f_n$ for $n>2$ can be written as follows
\begin{eqnarray}
f_n&=&\left.\frac{\partial^n F(\mu)}{\partial\mu^n}\right|_{\mu=0}=\left.\frac{\partial}{\partial\mu} \left(\frac{\partial^{(n-1)} F(\mu)}{\partial\mu^{(n-1)}}\right)\right|_{\mu=0}\label{afS6}\\
&=&\dots=x_n \langle N\rangle^{(n-1)b-(n-2)}|_{\mu=0}\nonumber,
\end{eqnarray}
where
\begin{equation}\label{afS7}
x_n=(-1)^na^{n-1}\prod_{i=2}^n[(n-2)b-(n-3)],
\end{equation}
and $\langle N\rangle$ is given by Eq.~(\ref{meanN}).

Eqs.~(\ref{afS6}) and~(\ref{afS7}) have been used to prepare Fig.~\ref{fig4}.

\subsection{Number of states characterizing the Poisson distribution}\label{AppendixPoisson}

The Poisson distribution,
\begin{equation}\label{P1}
P(N;\mu)=\frac{e^{-\langle N\rangle}\langle N\rangle^N}{N!},
\end{equation}
fulfills Taylor's power law (\ref{TaylorLaw}) with parameters $a=b=1$. The mean value (\ref{meanN}) and the free energy (\ref{FreeEnergy}) corresponding with this distribution are given by
\begin{equation}\label{P3}
\langle N\rangle=Xe^{-\mu}
\end{equation}
and
\begin{equation}\label{P4}
F(\mu)=\langle N\rangle+Y=Xe^{-\mu}+Y,
\end{equation}
where $X$ and $Y$ represent integration constants. In the following we calculate the number of states $g(N)$ characterizing the Poisson distribution in a twofold manner: first, taking advantage that we know the closed formula for the distribution (\ref{P1}), and second, using Eqs.~(\ref{new1}),~(\ref{new2}), and~(\ref{SPTDOS2}). In the second derivation one does not need to know the closed formula for the distribution (it is only known in a few cases, including the considered case of the Poisson distribution). The whole information about the system is taken from the free energy $F(\mu)$.

Thus, putting Eq.~(\ref{P3}) into Eq.~(\ref{P1}) one gets the following form of the considered distribution
\begin{equation}\label{P5}
P(N;\mu)=\frac{e^{-Xe^{-\mu}}(Xe^{-\mu})^N}{N!}.
\end{equation}
Similarly, inserting Eq.~(\ref{P4}) into Eq.~(\ref{defDOS}) one gets the equivalent formula for the distribution
\begin{equation}\label{P6}
P(N;\mu)=g(N)\frac{e^{-\mu N}}{e^{Xe^{-\mu}+Y}}.
\end{equation}
Comparing Eqs.~(\ref{P5}) and~(\ref{P6}) one finds that $Y=0$ and $g(N)$ corresponding to the Poisson distribution is given by
\begin{equation}\label{P7}
g(N)=\frac{X^N}{N!}.
\end{equation}
The expression (\ref{P7}) for the number of states has a very simple interpretation. The numerator of $g(N)$ corresponds to the number of $N$-variations (each of size $N$) from a set of size $X$. It means that the Poisson distribution arises in such systems, where each of $N$ elements (birds, cars, etc.) can be found in one of $X$ states, regardless of state of other elements. The denominator, $N!$, of $g(N)$ automatically assures us that elements of the considered system are indistinguishable.

In the following, we show that the same formula for the number of states characterizing the Poisson distribution can be directly obtained form Eqs.~(\ref{new1})-(\ref{SPTDOS2}). To use the mentioned equations, one has to find the coefficients $f_n$ (\ref{fn}) in the MacLaurin  expansion of the free energy $F(\mu)$ (\ref{P4}). It is easy to see that the coefficients are given by
\begin{equation}\label{P8}
f_n=(-1)^nX.
\end{equation}
Inserting the coefficients into Eq.~(\ref{SPTDOS2}) one gets
\begin{equation}\label{P9}
\mathcal{G}(n)=\frac{e^{X}}{n!}\sum_{k=1}^nX^kS(n,k),
\end{equation}
where $S(n,k)=B_{n,k}(1,1,\dots,1)$ stands for the Stirling number of the second kind. To derive Eq.~(\ref{P9}) we have make use of the well-known properties of Bell polynomials \cite{Bellinfo2}:
the first one,
\begin{equation}\label{P10}
B_n(x_1,x_2,\dots,x_n)=\sum_{k=1}^nB_{n,k}(x_1,x_2,\dots,x_{n-k+1}),
\end{equation}
stating that the $n$th complete Bell polynomial, $B_n$, is a sum of partial Bell polynomials, $B_{n,k}$, and the second one,
\begin{equation}\label{P11}
B_{n,k}(abx_1,ab^2x_2,\dots)=a^kb^nB_{n,k}(x_1,x_2,\dots),
\end{equation}
resulting directly from definition of $B_{n,k}$,
\begin{equation}\label{P12}
B_{n,k}(x_1,x_2,\dots)=\sum\frac{n!}{c_1!c_2!\dots(1!)^{c_1}(2!)^{c_2}\dots}x_1^{c_1}x_2^{c_2}\dots,
\end{equation}
where the summation takes place over all integers $c_1,c_2,c_3,\dots\geq 0$, such that
$c_1+2c_2+3c_3+\dots=n$ and $c_1+c_2+c_3+\dots=k$.

The sum on the right hand side of Eq.~(\ref{P9}) can be recognized as the right hand side of the so-called   Dobi\'{n}ski formula \footnote{E. W. Weisstein, {\it Dobi\'{n}ski's formula}. From MathWorld - A Wolfram Web Resource. http://mathworld.wolfram.com/DobinskisFormula.html},
\begin{equation}\label{P13}
\sum_{N=0}^\infty N^n\frac{e^{-X}X^N}{N!}=\sum_{k=1}^nX^kS(n,k),
\end{equation}
that gives the $n$th moment of the Poisson distribution with expected value $X$. Now, putting Eq.~(\ref{P13}) into Eq.~(\ref{P9}), and then comparing the resulting formula with Eq.~(\ref{new1}) one finds the number of states $g(N)$ given by the same formula, Eq.~(\ref{P7}), as the one derived at the beginning of this subsection.

\subsection{Fitting of experimental frequency distributions in Fig.~\ref{fig2}}\label{fitting}

Theoretical frequency distributions, $P(N)$, describing systems with Taylor's fluctuation scaling have the following form (cf.~Eqs.~(\ref{defDOS}) and~(\ref{apf1}))
\begin{equation}\label{EqS9}
P(N)=g(N)\frac{e^{-\mu N}}{e^{F(\mu)}}.
\end{equation}
To fit the experimental data with the formula, one has to find the correct value of the parameter, $\mu$, and also use the proper expressions for the free energy, $F(\mu)$, and the number of states, $g(N)$. The value of $\mu$ can be calculated from the formula for the first moment of the distribution $\langle N\rangle$ (\ref{meanN}). The function, $F(\mu)$, is given by (\ref{FreeEnergy}). A problem, however, arises with $g(N)$ underlying the considered systems. The closed form expression for $g(N)$ only exists in a few cases of the Taylor's parameter, $b$. Of course, in the limit of $N\gg 1$, one can approximate $g(N)$ with Eq.~(\ref{DOSfinal}), but the numerical computation of the corresponding Bell polynomials, $B_N(f_1,f_2,\dots,f_N)$, is extremely challenging and time-consuming. Therefore, in order to fit experimental frequency distributions in Fig.~\ref{fig2} with Eq.~(\ref{EqS9}) we have applied a numerical procedure whose details are exposed below.

Let us note that, although the formula for the frequency distribution, $P(N)$ (\ref{EqS9}), does not have a closed-form expression for arbitrary values of $b\geq 1$, the closed-form expression for generating function, $\mathbf{P}(s)$, of this distribution does exist:
\begin{equation}\label{EqS10}
\mathbf{P}(s)=\sum_{N=0}^\infty P(N)s^N=e^{F(\mu-\ln s)-F(\mu)}.
\end{equation}
Thus, the probability distribution, $P(N)$, may be obtained from the series expansion of $\mathbf{P}(s)$ at $s=0$, i.e.,
\begin{equation}\label{EqS11}
P(N)=\frac{1}{N!}\left[\frac{d^{(N)}
\mathbf{P}(s)}{ds^N}\right]_{s=0}.
\end{equation}
Unfortunately, here, due to the fact that $\ln s$ and its consecutive derivatives diverge at $s=0$ standard numerical procedures do not cope with Eq.~(\ref{EqS11}). To overcome this problem, we have expanded the logarithm in (\ref{EqS10}) into a power series up to the second order, i.e., $\ln s\simeq (s-1)-\frac{(s-1)^2}{2}$. Having a new generating function,
\begin{equation}\label{EqS12}
\widetilde{\mathbf{P}}(s)=\sum_{N=0}^\infty P(N)s^N=e^{F\left(\mu-(s-1)+\frac{(s-1)^2}{2}\right)-F(\mu)},
\end{equation}
we were able to calculate its consecutive derivatives at $s=0$ and also the corresponding approximated frequency distribution, $\widetilde{P}(N)$ (\ref{EqS11}).

Fig.~\ref{fig2} in the main body of the article presents four experimental  frequency distributions fitted with the approximated probability distribution, $\widetilde{P}(N)$. In the Figure, a direct fitting of the distribution describing stock market dynamics has been unsuccessful due to practical constraints related to impossibility of calculating derivatives of the desired order, $10^7$, for (\ref{EqS11}). Nevertheless, a theoretical frequency distribution characterizing NYSE has been obtained indirectly thanks to the scale-free character of Taylor's power law. The procedure of the indirect fitting consisted of rescaling the parameter, $N$, with the factor, $z$, according to $N\rightarrow N/z$. Taylor's law describing the rescaled systems has the following form
\begin{equation}
\frac{\sigma_N^2}{z^2}=a^*\left(\frac{\langle N\rangle}{z}\right)^b,
\end{equation}
providing us with the new value of Taylor's parameter,
\begin{equation}
a^*=az^{b-2},
\end{equation}
which characterizes the frequency distribution in (\ref{EqS11}) describing the rescaled parameter, $N/z$.

\section{Supplementary discussion}

\subsection{Key factor analysis}

The first remark relates to what we call key factor analysis. Let us note that, although Eq.~(\ref{DOSfinal}) is the exact theoretical result, which helps to understand the meaning of the number of states, in reality it may not be the simplest way to calculate the NoS. For example, in the case of three-dimensional, free electron $g(E)$ describes the number of states that are available to be occupied when the electron has energy $E$ and is given by $g(E)\propto\sqrt{E}$. The relation can be simply derived form the formula for the energy $E(p_x,p_y,p_z)= (p_x^2+p_y^2+p_z^2)/2m$, which depends only on momentum components, $p_x$, $p_y$, and $p_z$. The constant energy surface corresponds to the sphere of radius, $\sqrt{p_x^2+p_y^2+p_z^2}\propto \sqrt{E}$. Therefore, the number of states with energy $\leq E$ is given by $\Gamma(E)\propto E^{3/2}$, and the number of states with energy $E$ is $g(E)=d\Gamma(E)/dE\propto\sqrt{E}$.

The above example shows that calculation of the number of states, $g(N)$, underlying Taylor's power law should start by specifying the key factors accounting for the quantity $N$ in the considered systems (in the case of electron, the key factors are momentum components). The idea behind our theoretical approach is that a set of such factors exists and significantly contributes to the variation of $N$. An analogous idea underlies the key factor analysis (or the life table analysis) in population biology \cite{1959Morris}. Contrary to the classical approach to the problem proposed by biologists, however, our approach goes far beyond the analysis of the simplified key factor indices \cite{1996Royama}. We also would like to point out that the idea of key factors is particularly interesting in relation to stock market dynamics and dynamics of other man-made systems with Taylor's fluctuation scaling. Our derivations show that the systems possess a well-defined number of states and, therefore, a well-defined group of key factors.

\subsection{Dynamical processes leading to Taylor's law}

The second comment relates to dynamical processes leading to Taylor's law. It is important to understand that the proposed theoretical approach to the origins of the law provides static pictures of the phenomena. It does not explain details of the dynamical processes underlying the fluctuation scaling. The same is true for statistical mechanics, which reproduces static aspects of thermodynamics without dealing with dynamical processes such as the approach to equilibrium or dynamics in equilibrium. To investigate these points from the microscopic dynamics is indeed a very difficult and poorly understood problem (see e.g. \cite{1998PRLTasaki}). In most of cases, however, a rather satisfactory understanding of these issues is possible due to simple methods borrowed from the theory of stochastic processes and computational statistical physics. For example, the detailed balance condition \cite{BookNewman} applied to the known microscopic models accounting for Taylor's law \cite{1977NatureTaylor,1982NatureAnderson,1983NatureTaylor,2004PRLMenezes2} may help us to understand the meaning of the macroscopic parameter, $\mu$ cf.~Eqs.~(4)~-~(6). In this sense, our approach does not cancel the previous models. It merely allows identification of models' important features that account for the law. Finally, let us note that the meaning of $1/\mu$ is similar to the meaning of thermodynamic temperature, $T$. For physicists, it has taken a long time to understand the meaning of temperature and entropy. We believe that our approach will help to introduce and understand these concepts in other areas of science.


\end{document}